# Ultra-Wide Bandgap Gallium Oxide Films:
# UV-Luminescence and Phonon Dynamics at Extreme Temperatures

Isiaka Lukman[1], Matthew D. McCluskey[2], and Leah Bergman[1]

[1]*Department of Physics, University of Idaho, Moscow ID 83844 USA*

[2]*Department of Physics and Astronomy, Washington State University, Pullman WA 99164 USA*

Corresponding author: Leah Bergman, Lbergman@uidaho.edu

## Abstract

$\beta$-$Ga_2O_3$ is a semiconductor with bandgap in the deep-UV ~ 5 eV. Due to its strong phonon-hole coupling, holes are self-trapped inhibiting bandgap luminescence at the deep-UV. In contrast, the self-trapped holes (STH) can exhibit a strong luminescence at ~3.5 eV. This research addresses the thermal response of the STH photoluminescence (PL), and the role of phonon interactions at temperatures 77 K - 622 K in nanocrystalline films. It was found that the PL intensity strongly diminishes as a function of increasing temperature with activation energy ~ 72 meV. A study of the Raman modes revealed that the intensity of the high frequency modes of the $Ga_IO_4$ site decrease with temperature, implying a phonon annihilation process. These modes, which have comparable energy to the STH activation energy, thus can couple to the STH and transition them from a radiative to a non-radiative regime in accordance with the configurational coordinate model at the strong phonon coupling limit. The significantly broad Gaussian linewidth of the PL is also a manifestation of a strong STH-phonon coupling. Furthermore, the peak position of the STH exhibited a negligible temperature response, in contrast to the redshift of ~ 220 meV of the bandedge of the film. In contrast to the intensity behavior of the high frequency Raman modes, the low frequency ones were found to follow the thermal Bose-Einstein phonon population.

β-Ga$_2$O$_3$ is a semiconductor with ultra-wide bandgap in the deep UV range of ~ 5 eV [1]. Ga$_2$O$_3$ can crystalize in four distinct phases, among them the β is the most thermodynamically stable up to its melting point of ~ 1800 C [2-3]. The superb material properties of β-Ga$_2$O$_3$ enable its technological applications in UV-optoelectronic devices and high-temperature high-power sensors and photodetectors [4-8]. Moreover, due to its relatively low-phonon energies, Ga$_2$O$_3$-based low-energy phonon (LEP) wideband optical windows have been realized [5].

One of the most interesting crystal dynamics properties of β-Ga$_2$O$_3$ is its self-trapped hole (STH) that inhibits bandgap luminescence at the deep UV range ~ 5 eV. In their theoretical study, Varley et al. predicted that due to a large crystal distortion, holes become localized at the oxygen 2p orbitals [1]. As a result, the holes become highly immobile, which may be a deterrent to achieving *p*-type β-Ga$_2$O$_3$ and moreover impede the deep UV photoluminescence (PL) due to free e-h pair recombination at the bandgap energy range. In contrast, when a free electron from the conduction band, or an exciton, recombines with STH, a characteristic light emission at the near UV ~ 3 V is expected [1, 9-10], which can be very efficient even at room temperature [9].

As the luminescence of the STH is an inherent property of β-Ga$_2$O$_3$, and most often is the only dominant light emission, addressing its properties is of merit contributing to fundamental knowledge as well as to the technological applications of this material. This research focuses on the temperature response of the STH PL and phonon dynamics at 77 K – 622 K that addresses their fundamental properties and may provide vital knowledge for the thermal management of devices operating at extreme temperatures. The nanocrystalline morphology of the β-Ga$_2$O$_3$ sample resulted in the observation of most of the Raman modes, and in conjunction with the highly resolved PL spectra, enabled a detailed study of the phonon and STH properties at a wide temperature range.

In this study, β-Ga$_2$O$_3$ thin films, ~ 300 nm, were grown on quartz and silicon substrates via magnetron radio frequency (RF) magnetron sputtering at 400 C. After growth, the films were annealed at 1100 C in ambient atmosphere for 3 hours. Further details on growth can be found in reference 9. The photoluminescence and Raman experiments were acquired employing a high-resolution Jobin–Yvon T64000 micro-Raman and PL system optimized for the UV range. In conjunction with the PL system, a laser at 244 nm (5.1 eV) was used in order to achieve above band-gap excitation. The transmission study was carried via an Agilent Cary 300 spectrophotometer operating in a double beam mode. The PL, Raman, and transmission spectroscopes were studied at a temperature range of 77 K to 622 K using an Instec 621V microcell which has a UV compatible window.

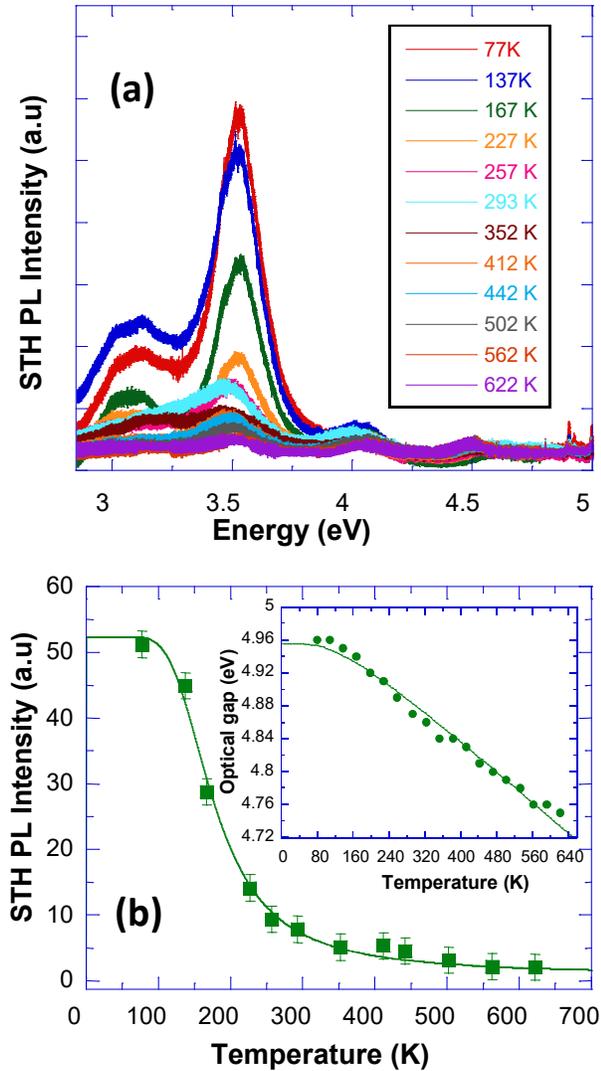

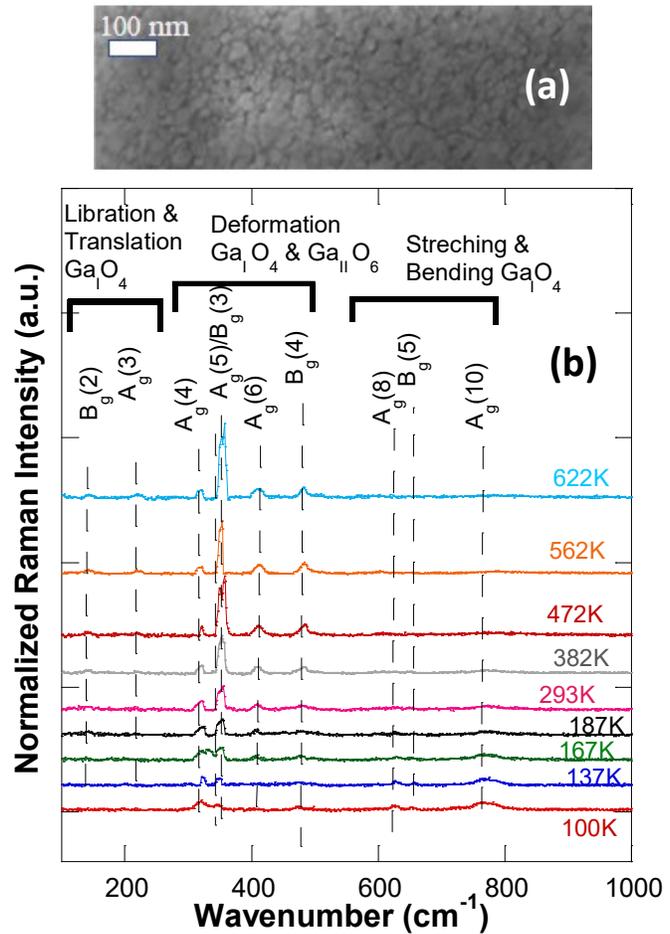

**Fig. 2** An SEM image of the β-Ga₂O₃ film showing its micro-crystalline morphology **(a)**, and Raman scattering spectra as a function of temperature **(b)**.

**Fig. 1** The highly resolved PL spectra as a function of temperature that spans a large range of energies at and below the bandgap of β-Ga₂O₃ film. The temperature range is 77 K to 622 K **(a)**. The intensity of the STH as a function of temperature: experimental data (points) and a line fit using Equation 1. The inset to the figure is the thermal response of the optical gap energy of the film **(b).**

Figure 1(a) shows the highly resolved PL of the film at a wide energy range that spans the near UV ~ 2.85 eV up to the deep UV ~ 5 eV which is the expected bandgap of β-Ga$_2$O$_3$. The spectra were taken at temperatures of 77 K to 622 K. The main peak at ~ 3.5 eV is assigned to the STH. The STH properties were previously studied by us and by other groups, via experiments as well as theory, where peak energies at 3.1 to 3.6 were reported [1, 9-11]. Also present in the spectra is a weaker peak ~ 3 eV attributed to donor acceptor transitions [9-10]. The key point is that the PL intensity of the STH is a strong function of temperature, implying that nonradiative mechanisms compete with the PL.

The behavior of the intensity, I(T), of the STH may be studied via the efficiency model of Equation 1 [12] :

$$I(T) = \frac{I_o}{\left(1+A*\exp(E_a/k_BT)\right)} \qquad (1)$$

where $I_o$ is the intensity at absolute temperature, $E_a$ is the activation energy needed for the PL to follow a nonradiative route, $A$ is a constant, and $k_B$ is Boltzmann's constant. Figure 1(b) presents the experimental data points of the intensity as a function of temperature at the range of 77 K to 622 K, while the line is a fit to Equation 1. The analysis yielded an activation energy ~ 71.4 meV. A similar activation energy, 72.1 eV, was previously reported for single crystals using X-ray excitation experiments and was attributed to the thermal quenching of STH due to migration to $V_{Ga}^{3-}$ centers [11]. The $V_{Ga}^{3-}$ were identified via a weak and broad luminescence peak centered at ~ 2.82 eV to 2.61 eV depending on the temperature [11]. Additionally, an activation energy of 200 meV was found by deep level transient spectroscopy study and was attributed to the transition of STH to mobile holes in the valance band [13]. However, theoretical calculations predicted that the self-trapping energy for STH of β-Ga₂O₃ is much higher than the 70 meV and 200 meV discussed above; a value of 530 meV was calculated by [1], implying that an activation energy of that value is needed for the hole in the STH state to transition to the valence band, become mobile, and interact with traps in the crystal. The issue of the STH dynamics is thus yet an open question, and in principle several routes that result in the quenching of the PL may coexist. As will be discussed in the following paragraphs, we found that phonon interactions may provide a vital nonradiative mechanism to the STH PL.

The optical properties of our β-Ga₂O₃ samples were previously studied by us and the thermal response of the optical gap is presented in the inset of Figure 1(b) [9]. As can be seen in the inset, the optical gap ranges from 4.96 eV to 4.75 depending on the temperature, and at room temperature is at 4.85 eV [9]. The PL spectra in Figure 1(a) shows a significant quenching of the STH PL; however, no change in PL intensity is observed in the deep UV range corresponding to the optical gap energy of the film. Specifically, if the activation energy found here were to correspond to the energy needed to transition holes into the valence band, a free electron-hole recombination should occur at the higher temperature regime. This should result in an increased PL at ~ 4.8 eV which is not observed in the spectra.

A mechanism that can explain the strong PL quenching of the STH is a nonradiative process that involves phonons. The rationale for considering this mechanism is that β-$Ga_2O_3$ has a very strong phonon coupling due the large crystal distortion [1]. In order to gain insight into the effect of phonons on the PL of the STH, the temperature response of the Raman modes was investigated. In general, β-$Ga_2O_3$ has 15 Raman active modes: the ones that are observed in a spectrum depend on the Raman selection rules that take in consideration the polarization of the incoming and outgoing light and a specific crystal plane [14].

As can be seen in the scanning electron microscope (SEM) image presented in Figure 2(a), the film has granular morphology with randomly oriented planes. This type of morphology is conducive to what is known as the breaking of the Raman selection rules which results in Raman scattering of multiple modes. Figure 2(b) presents the Raman spectra of the β-$Ga_2O_3$ as a function of temperature at a range 100 K to 622 K. The Raman mode frequencies of our study concur with the ones observed previously for single crystals and ceramics β-$Ga_2O_3$ [14-15]. The Raman modes of β-$Ga_2O_3$ can be categorized into three groups: (i) low frequency modes due to libration and translation of the $Ga_IO_4$ chains, (ii) mid frequency modes due to deformation of the $Ga_IO_4$ and $Ga_{II}O_6$ sites, and (iii) high frequency modes due to stretching and bending of $Ga_IO_4$ sites. In the β-$Ga_2O_3$ crystal structure, $Ga_IO_4$ is the tetrahedral site while the $Ga_{II}O_6$ is the octahedral site. For this study we focused on groups (ii) and (iii) that will be referred to as the low and high energy modes, respectively.

An inspection of the temperature response of the Raman mode intensity indicates that the modes can be separated into two sets: one that exhibits a decrease in intensity as a function of increasing temperature, and the other that exhibits an increase in intensity with increasing temperature as can be seen in Figures 2(b) and 3(a,b). These findings suggest that the first set of phonons, mainly the $A_g(8)$ and $A_g(10)$, may be absorbed and annihilated as a function of raising temperature. Specifically, as can be seen in Figure 3(a), the $A_g(8)$ and the $A_g(10)$ modes, that are due to the stretching and bending of the $Ga_IO_4$ site, which is also the crystallographic location of the STH, have phonon energies ~ 78 meV and 95 meV respectively. These phonon energies, up to experimental error, are comparable to the activation energy of the STH ~ 72 meV, in particular the energy of the $A_g(8)$ mode. The above analysis may indicate that this group of phonons are being absorbed by the STH and energetically enable its non-radiative transition.

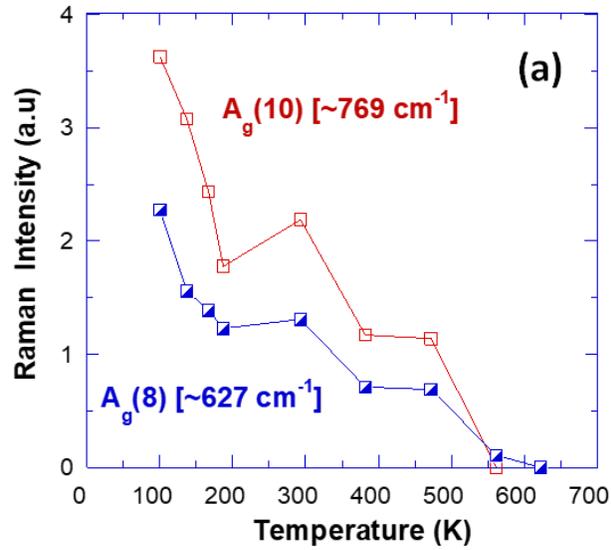
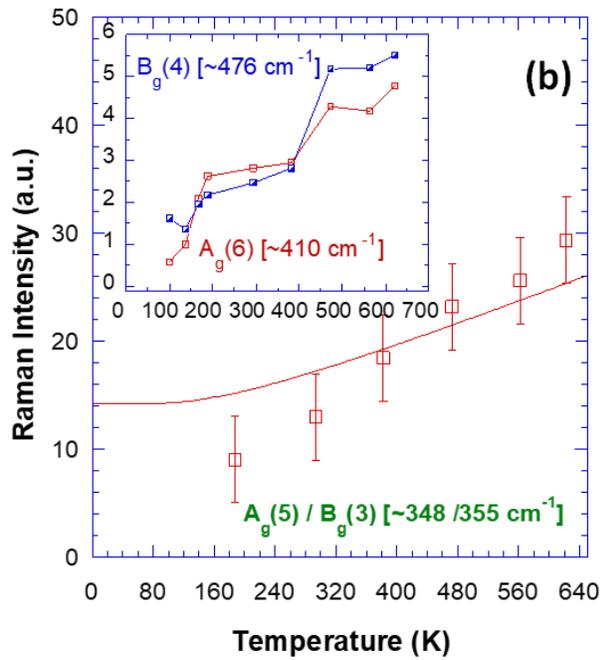
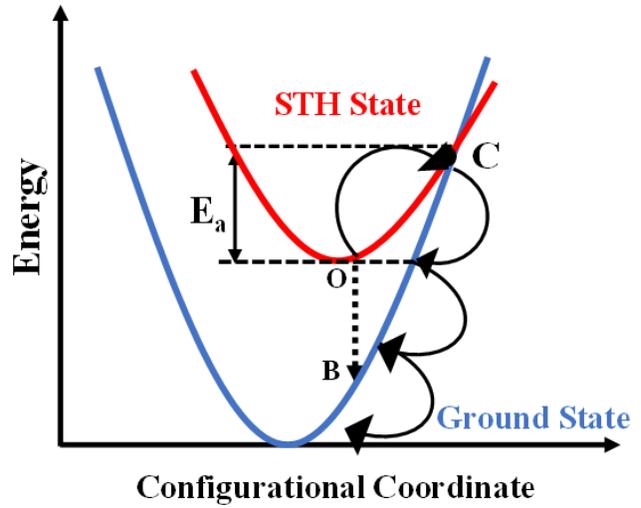
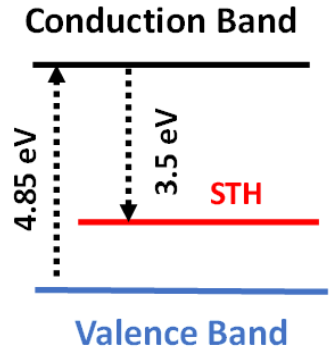

**Fig. 4** A diagram for the STH configurational coordinate model, in the case of strong phonon coupling, and its representative energy band diagram, describing the energetics of the STH. The experimental values are from our results.

**Fig. 3** The thermal response of the Raman intensity for the high frequency phonon modes $A_g(10)$ at 769 cm$^{-1}$, and $A_g(8)$ at ~ 627 cm$^{-1}$ **(a)**. The thermal response of the low frequency mode $A_g(5)/B_g(3)$ doublet (348/355 cm$^{-1}$), and a fit of the Bose Einstein phonon population model of Equation 2. The inset is the thermal behavior of the other low frequency modes $B_g(4)$ at 476 cm$^{-1}$, and that of $A_g(6)$ at ~ 410 cm$^{-1}$ **(b)**.

In consideration of the above results, we propose a phonon mediated nonradiative route which is based on the configurational coordinate model for a non-radiative process [12, 16-18]. As was mentioned previously, the rationale for that choice stems from the strong crystal distortion associated with the STH of β-$Ga_2O_3$. Figure 4 shows a diagram of the configurational coordinate model for the ground state of the bulk and the STH excited state, for which, due to the strong lattice distortion, their minimum points are displaced, and a crossover point $C$ is formed. At moderate temperatures, the STH luminesces (point $O$ to $B$). When the STH is thermally excited with $E_a$ from point $O$ to $C$ via interaction with $A_g(10)$ and $A_g(8)$ phonons, it is transitioned to a non-radiative regime at point $C$, and the luminescence intensity diminishes as a result. The nonradiative relaxation then takes place to the ground state via multiple phonon scattering. The phonon dispersion curve of β-$Ga_2O_3$ is highly dense [6, 19], and therefore routes to the ground state may be feasible.

In contrast to the behavior of the high energy phonons discussed above, the low energy group of $A_g(6)$, $B_g(4)$, and specifically the $A_g(5)/B_g(3)$ doublet show an intensity increase with temperature, as can be seen in Figures 2(b) and 3(b). Due to the more pronounced intensity increase of these phonons, with energies ~ 44 meV, they were used in the following analysis. In the harmonic approximation, the Raman intensity of a specific mode can be related to a thermal factor of the form of Bose-Einstein phonon population that increases with temperature [20]:

$$\frac{I(T)}{I_o} = \frac{1}{\left[exp\left(\frac{0.044}{8.617*10^{-5}*T}\right)\right]-1} + 1 \qquad (2)$$

In this relation 0.044 eV is the phonon frequency, $I_o$ is the intensity as the temperature approaches zero, and 8.617 x$10^{-5}$ is Boltzmann constant in units of eV/K. As can be seen in Figure 3(b), the intensity increase of the $A_g(5)/B_g(3)$ approximately follows the above model calculation at the temperature range of 167 K – 622 K (at lower temperatures the intensities were too low for meaningful analysis).

A further study of the optical properties of the STH found that there is an insignificant peak shift at the wide temperature range of 77 K to 622 K. This is in contrast with the 220 meV shift of optical gap of our β-Ga$_2$O$_3$ microcrystalline films, as can be seen in the inset to Figure 1(b) [9]. A weak temperature dependent PL peak position is characteristic of a deep level defect with strong electron-phonon coupling [21] as is the case for β-Ga$_2$O$_3$. A weak temperature response of the STH peak position was also observed previously in β-Ga$_2$O$_3$ single crystals [11], implying this phenomenon is a fundamental property of this material. A manifestation of the strong STH-phonon coupling in β-Ga$_2$O$_3$ should be reflected in a Gaussian shaped PL with a very broad linewidth. The PL spectra of β-Ga$_2$O$_3$ and ZnO films are presented in Figure 5. The PL linewidth of the STH is significantly broad ~ 216 meV: this contrasts with the linewidth, ~ 30 meV, of the band edge PL of the ZnO microcrystalline film and nanocrystals [22-23]. Moreover, as can be seen in Figure 5, the line shape of the STH is well approximated by a Gaussian function (deconvoluted from the lower energy PL due to donor acceptor emission). Very broad Gaussian emission lines are characteristic of strong phonon-optical center coupling, as in the case of β-Ga$_2$O$_3$. In such vibronic interaction, the luminescent center can absorb and emit phonons, resulting in broad sidebands by what is known as the zero-phonon line (ZPL) [18, 24-25].

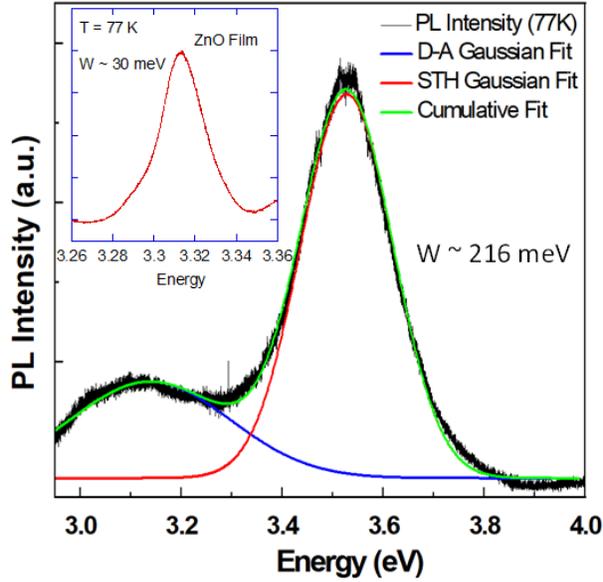

**Fig. 5** The PL of the STH and that of the band edge PL of ZnO microcrystalline film (inset) at 77 K. The linewidths are 216 meV and 30 meV, respectively. The STH exhibits a Gaussian lineshape.

In summary, the luminescence and Raman mode behavior of β-Ga$_2$O$_3$ films were studied at a wide temperature range of 77 K – 622 K. The highly resolved UV PL spectra enabled a meaningful analysis of the temperature response of the STH light emission at ~ 3.5 eV. It was found that the PL intensity exhibits a strong quenching with raising temperature, with activation energy of ~ 72 meV. The PL spectra at the ~ 5 eV range did not show any significant increase in the PL intensity at the band edge, implying that the STH did not transition into the valence band. The high-energy Raman modes, which have comparable Raman energy to that of the STH activation energy, were discussed in terms of the configurational coordinate model in which the high energy phonons activate the STH into a non-radiative regime. In contrast, the low energy mode follows a Bose-Einstein type phonon population. Moreover, the STH energy peak has insignificant response to temperature, a behavior which is similar to that of a deep level defect. The STH PL exhibits a Gaussian shape with extreme broadening, both of which are manifestations of strong STH-phonon coupling.

## Acknowledgements

This research was supported by the U.S. Department of Energy, Office of Basic Energy Sciences, Division of Materials Science and Engineering under award DE-FG02-07ER46386.

## References

**1.** J. B. Varley, A. Janotti, C. Franchini, and C. G. Van de Walle, Phys. Rev. B **85** (8), 081109 (2012).

**2.** Y. Yao, S. Okur, L. A. M. Lyle, G. S. Tompa, T. Salagaj, N. Sbrockey, R. F. Davis, and L. M. Porter, Mater. Res. Lett. **6** (5), 268 (2018).

**3.** S. Lee, K. Akaiwa, and S. Fujita, Phys. Status Solidi C **10** (11), 1592 (2013).

**4.** B. R. Tak, M. Garg, S. Dewan, C. G. Torres-Castanedo, K.-H. Li, V. Gupta, X. Li, and R. Singh, J. Appl. Phys. **125,** 144501 (2019).

**5.** K. Yoshimoto, A. Masuno, M. Ueda, H. Inoue, H. Yamamoto, and T. Kawashima, Sci. Rep. **7** (1), 45600 (2017).

**6.** K. A. Mengle and E. Kioupakis, AIP Adv. **9**, 015313 (2019).

**7.** S. J. Pearton, J. Yang, P. H. Cary, F. Ren, J. Kim, M. J. Tadjer, and M. A. Mastro, Appl. Phys. Rev. **5,** 011301 (2018).

**8.** M. Bartic, C.-I. Baban, H. Suzuki, M. Ogita, and M. Isai, J. Am. Ceram. Soc. **90**(9), 2879 (2007).

**9.** D. Thapa, J. Lapp, I. Lukman, and L. Bergman, AIP Advances **11**, 125022 (2021).

10. T. Onuma, Y. Nakata, K. Sasaki, T. Masui, T. Yamaguchi, T. Honda, A. Kuramata, S. Yamakoshi, and M. Higashiwaki, J. Appl. Phys. **124**, 075103 (2018).

11. H. Tang, N. He, Z. Zhu, M. Gu, B. Liu, J. Xu, M. Xu, L. Chen, J. Liu, X. Ouyang, Appl. Phys. Lett. **115,** 071904 (2019).

12. "Luminescence in Crystals", by D. Curie (John Wiley & Sons Inc, New York, 1963).

13. A.Y. Polyakov, N.B. Smirnov, I.V. Shchemerov, S.J. Pearton, F. Ren, A.V. Chernykh, P.B. Lagov, and T.V. Kulevoy, APL Materials **6**, 096102 (2018).

14. T. Onuma, S. Fujioka, T. Yamaguchi, Y. Itoh, M. Higashiwaki, K. Sasaki, T. Masui, and T. Honda, Journal of Crystal Growth **401**, 330 (2014).

15. D. Dohy, G. Lucazeau, and A. Revcolevschi, Journal of Solid-State Chemistry **45,** 180 (1982).

16. D.L. Dexter, C.C. Klick, and G.A. Russell, Phys. Rev. **100,** 603 (1955).

17. M.A. Reshchikov, J. Appl. Phys. **115,** 012010 (2014).

18. K.M. McCall, C.C. Stoumpos, S.S. Kostina, M.G. Kanatzidis, B.W. Wessels, Chem. Mater. **29**, 4129 (2017).

19. K. Ghosh and U. Singisetti, Appl. Phys. Lett. **109**, 072102 (2016).

20. H. Wang, F.D. Medina, Y.D. Zhou, and Q.N. Zhang, Phys. Rev. B. **45**, 10356 (1992).

21. M. A. Reshchikov, F. Shahedipour, R.Y. Korotkov, B.W. Wessels, and M.P. Ulmer, J. Appl. Phys. **87,** 3351 (2000).

22. D. Thapa, J. Huso, J.L Morrison, CD. Corolewski, M.D. McCluskey, and L. Bergman, Optical Materials **58,** 382 (2016).

23. J.L. Morrison, J. Huso, H. Hoeck, E. Casey, J. Mitchell, L. Bergman, and M.G. Norton, J. Appl. Phys. **104**, 123519 (2008).

- 11 -
**24.** "Optical Spectroscopy of Inorganic Solids", by B. Henderson and G.F. Imbusch, (Oxford University Press, New York, 1989).

**25.** A. Alkauskas, M.D. McCluskey, and C.G. Van de Walle, J. Appl. Phys. **119**, 181101 (2016).